\documentstyle[epsf]{elsart}

\begin{document}
\begin{frontmatter}
\title{Microscopic shell-model description of the exotic
nucleus $^{16}$C}

\author[CNS,Tokyo]{S. Fujii},
\author[CNS,Senshu]{T. Mizusaki},
\author[CNS,Tokyo,RIKEN]{T. Otsuka},
\author[Hosei]{T. Sebe},
\author[JSF]{A. Arima}

\address[CNS]
{Center for Nuclear Study (CNS), University of Tokyo,
Wako Campus of RIKEN, Wako 351-0198, Japan}
\address[Tokyo]
{Department of Physics,
University of Tokyo, Tokyo 113-0033, Japan}
\address[Senshu]
{Institute of Natural Sciences, Senshu University,
Tokyo 101-8425, Japan}
\address[RIKEN]
{RIKEN, Wako 351-0198, Japan}
\address[Hosei]
{Faculty of Engineering, Hosei University, Koganei 184-8584, Japan}
\address[JSF]
{Science Museum, Japan Science Foundation, Tokyo 102-0091, Japan}

\begin{abstract}
The structure of the neutron-rich carbon nucleus $^{16}$C is
described by introducing a new microscopic shell model of no-core type.
The model space is composed of the 0$s$, 0$p$, 1$s$0$d$, and 1$p$0$f$
shells.
The effective interaction is microscopically
derived from the CD-Bonn potential and the Coulomb force
through a unitary transformation theory.
Calculated low-lying energy levels of $^{16}$C agree well with
the experimental values.
The $B(E2; 2_{1}^{+} \rightarrow 0_{1}^{+}$) value is calculated
with the bare charges.
The anomalously hindered $B(E2)$ value for $^{16}$C, measured recently,
is well reproduced.
\end{abstract}
\end{frontmatter}

The study of nuclear structure far from the stability line is being
developed with great interests from both the experimental and
theoretical sides.
One of the recent exciting phenomena on these lines is the anomalously
hindered $E2$ transition strength between the first $2^{+}$ and the
ground $0^{+}$ states in $^{16}$C~\cite{Imai04}, while this transition
is rather strong in other nuclei.
The dominance of neutron excitation is also suggested for the same
transition~\cite{Elekes04,Ong06}.

In order to theoretically investigate such interesting properties,
several calculations have been performed.
For neutron-rich carbon isotopes with the mass number $A\geq 15$,
neutrons begin to fill the 1$s_{1/2}$ and 0$d_{5/2}$ orbits,
while the protons occupy mainly the orbits up to 0$p_{3/2}$.
Low-lying energy levels of such systems can be well described by
the conventional shell model using effective interactions for the
0$p$1$s$0$d$ shell~\cite{Warburton92,Maddalena01,Fujimoto03,Sagawa04}.
However, the $B(E2; 2_{1}^{+} \rightarrow 0_{1}^{+}$) values 
calculated by the shell model using the usual effective
charges are significantly larger than the experimental ones
though the systematic decrease from $^{14}$C to $^{16}$C is described.
A similar tendency of $B(E2; 2_{1}^{+} \rightarrow 0_{1}^{+}$)
is seen in the calculations using the bare
charges within the anti-symmetrized molecular dynamics (AMD)
framework~\cite{Thiamova04,Enyo05}.
The discrepancies between the calculations and the experiments
may be due to the fact that those models do not include certain exotic
features of unstable nuclei because of their bases with
some assumptions or adjustments in and near stable nuclei.

Given this situation, ab initio nuclear structure calculations
starting with the two- and three-body nuclear forces in free space
may help to describe correctly the nuclear structure
without any assumptions.
For this purpose, the Green's function Monte
Carlo (GFMC)~\cite{Pieper01,Pieper05} and the no-core shell model
(NCSM)~\cite{Navratil00,Navratil01} are promising methods up to
the mass number $A\simeq 12$.
As the mass number becomes larger, however, those methods may not be
practical because of a huge computation size.
It is, therefore, still of much importance to construct a model for
describing light nuclei starting with the bare nuclear force.

In this Letter, we shall show that
the $B$($E2; 2_{1}^{+} \rightarrow 0_{1}^{+}$) value as well as
low-lying excited levels of $^{16}$C can be well described by
a new shell-model framework being introduced.
The model space consists of the 0$s$, 0$p$, 1$s$0$d$, and 1$p$0$f$
shells.
Since the model space is of no-core type and is rather large as
compared to that of the conventional shell model,
we do not employ the effective charges but use the bare charges.
Although, in principle, a sort of effective operator is needed for
the calculation of
$B$($E2; 2_{1}^{+} \rightarrow 0_{1}^{+}$)~\cite{Stetcu05},
we use the bare operator for simplicity.
The effective two-body interaction is microscopically derived from
a bare nucleon-nucleon interaction through a unitary
transformation~\cite{Okubo54,Suzuki82} which is used in
the unitary-model-operator approach (UMOA)~\cite{Suzuki94,Fujii04}.

\begin{figure}[t]
  \epsfxsize = 12.5cm
  \centerline{\epsfbox{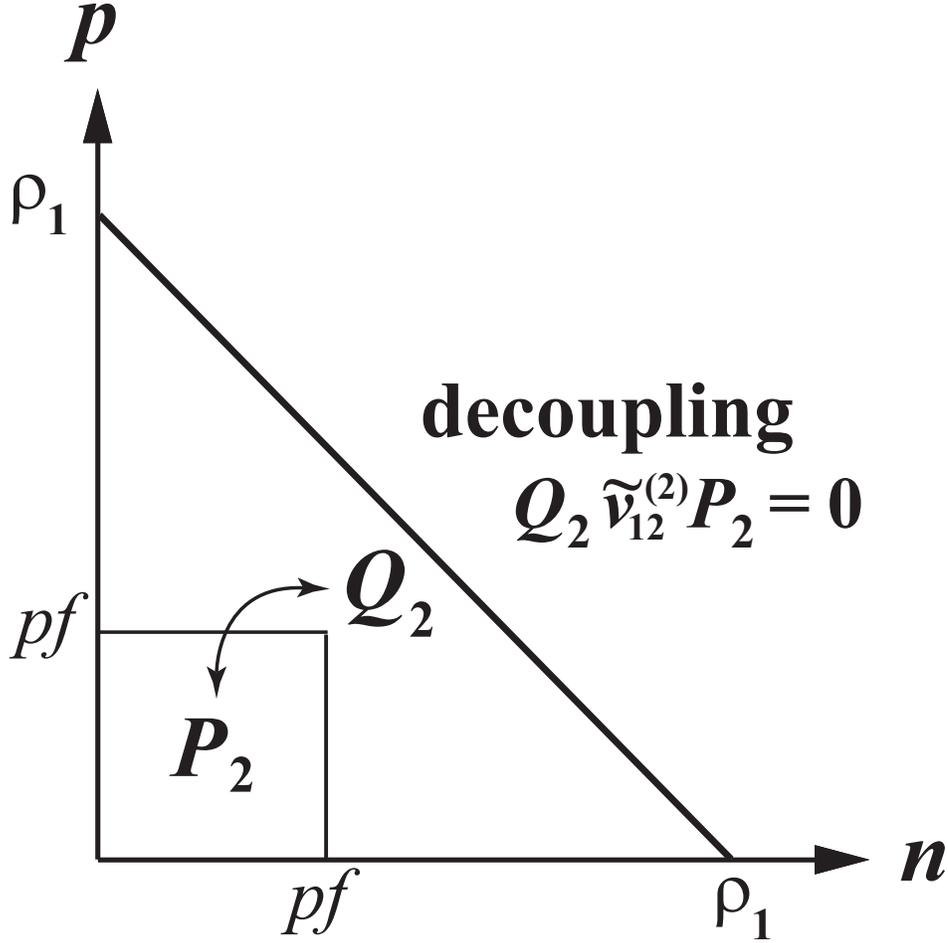}}
\caption{\label{fig:mspace} The model space $P_{2}$ and its complement
$Q_{2}$ consisting of the two-body states.
The axes represent nuclear shells.}
\end{figure}

Here we outline the method of deriving the effective interaction
for the present shell model.
Although the method is essentially the same as in the UMOA in
Ref.~\cite{Fujii04}, there are several differences regarding
the model space.
In Fig.~\ref{fig:mspace}, the model space $P_{2}$ and its complement
$Q_{2}$ for the neutron-proton channel are illustrated.
The other channels are also treated similarly.
First we derive the effective interaction $\tilde{v}^{(1)}_{12}$ in
the large model space $P_{1}(=P_{2}+Q_{2})$ which is specified by
a boundary number $\rho _{1}$ through the same procedure
as in Ref.~\cite{Fujii04}.
The number $\rho _{1}$ is defined as
$\rho _{1}=2n_{1}+l_{1}+2n_{2}+l_{2}$ with the harmonic-oscillator
(h.o.) quantum numbers \{$n_{1},l_{1}$\} and \{$n_{2},l_{2}$\}
of the two-body states.
If we diagonalize the effective Hamiltonian in the $P_{1}$ space with
a large number of shell-model many-body basis states,
it leads to the NCSM.
However, since the mass number in the present calculation is around 16,
it is difficult to obtain a converging result by enlarging the model
space $P_{1}$.
Hence, in order to derive an effective interaction suitable for
a smaller model space, we employ the procedure stated below.

The $P_{1}$ space is divided into the small model space $P_{2}$ and
the complementary space $Q_{2}$ as in Fig.~\ref{fig:mspace}.
By using the effective interaction $\tilde{v}^{(1)}_{12}$ in the
$P_{1}$ space where the short-range correlation is taken into account,
we derive the effective interaction $\tilde{v}^{(2)}_{12}$
in the $P_{2}$ space for the present shell-model calculation.
The $P_{2}$ space contains the orbits from the $0s$ up to the 1$p$0$f$
shell for both the proton and neutron.
The two-body effective interaction $\tilde{v}^{(2)}_{12}$ in
the $P_{2}$ space is determined by solving the decoupling
equation~\cite{Suzuki80}
\begin{equation}
\label{eq:decoupling}
Q_{2}\tilde{v}^{(2)}_{12}P_{2}=
Q_{2}\{ e^{-S_{12}}(h_{0}+\tilde{v}^{(1)}_{12})
e^{S_{12}}-h_{0}\} P_{2}=0,
\end{equation}
where $h_{0}$ is the one-body part of the two-body subsystem which
consists of the kinetic energy $t_{1}$ ($t_{2}$) and
the one-body potential $u_{1}^{(2)}$ ($u_{2}^{(2)}$) as
$h_{0}=t_{1}+u^{(2)}_{1}+t_{2}+u^{(2)}_{2}$.
The $u^{(2)}_{1}$ stands for a mean potential for the nucleon $1$
from the other nucleons which are assumed to occupy the orbits
according to the normal filling scheme for $^{14}$C.
The $u^{(2)}_{1}$ and $\tilde{v}^{(2)}_{12}$ are
self-consistently determined for the unperturbed ground-state
wave function of $^{14}$C by means of iteration.
This means that we derive the two-body effective interaction
$\tilde{v}^{(2)}_{12}$ which can describe $^{14}$C as a good
closed-shell nucleus.
This self-consistency is also used to determine the effective
interaction in the $P_{1}$ space.

It is known that, with the usual projection operators $P$ and $Q$,
the general solution $S_{12}$ for the decoupling equation is written
with the restrictive conditions $PS_{12}P=QS_{12}Q=0$ as
$S_{12}={\rm arctanh}(\omega-\omega^{\dagger})$,
where the operator $\omega$ satisfies
$\omega = Q\omega P$~\cite{Suzuki82}.
The unitary transformation $e^{S_{12}}$ is also expressed in terms of
the operator $\omega$ by the block form concerning $P$ and $Q$ as
\begin{eqnarray}
\label{eq:U_block}
e^{S_{12}}&=&
(1+\omega-\omega ^{\dagger})
(1+\omega \omega ^{\dagger} +\omega ^{\dagger}\omega )^{-1/2} \nonumber \\
&=&\left(
  \begin{array}{cc}
       P(1+\omega ^{\dagger}\omega)^{-1/2}P
    & -P\omega ^{\dagger}(1+\omega \omega ^{\dagger})^{-1/2}Q \nonumber \\
       Q\omega (1+\omega ^{\dagger}\omega )^{-1/2}P
    &  Q(1+\omega \omega ^{\dagger})^{-1/2}Q
  \end{array}
\right)\\
\end{eqnarray}
which agrees with the unitary transformation by \=Okubo~\cite{Okubo54}.

We note here that the model space $P_{2}$ is not completely separated
from the $Q_{2}$ space in energy but in the orbits.
Some two-body states in the $P_{2}$ space can be degenerate with those
in the $Q_{2}$ space in energy.
This degeneracy often causes difficulty with the convergence of
the self-consistent calculation.
This means that some matrix elements of $\tilde{v}^{(2)}_{12}$ in
a step are significantly different from those in another step.
As we enlarge the $P_{1}$ space by taking a large value of $\rho _{1}$,
this instability increases.
In the present study, we take $\rho _{1}=6$ as the maximum value for
obtaining the converged effective interaction.
In this case, the model space $P_{2}$ is bounded on the middle of the
diagonal line specified by $\rho _{1}$.
On the other hand, in the usual NCSM, one takes a sufficiently large
value of $\rho _{1}$ for obtaining the converged result and does not
perform the double unitary transformation for determining the effective
interaction in the small model space $P_{2}$ as in the present
work.
Therefore, the present approach may be regarded as a sort of
{\it model} to describe low-lying states in neutron-rich carbon
isotopes though that is still based on the bare nucleon-nucleon force
and is thus microscopic.
It is also noted that the self-consistent one-body potential
$u^{(2)}_{1}$ is introduced only in the $P_{2}$ space
for the convergence.
For the one-body potential in the $Q_{2}$ space in solving
the decoupling equation in Eq.~(\ref{eq:decoupling}),
we use the fixed one which is self-consistently determined with
the effective interaction in the large model space
$P_{1}(=P_{2}+Q_{2})$.

Next, we mention truncations in the actual shell-model calculation.
The nucleon excitations from the hole states of $^{14}$C are restricted
to up to two nucleons, and the excitations to the 1$p$0$f$ shell are
also up to two nucleons.
The three-or-more-body effective interactions are not taken into
account for simplicity though the many-body effective interactions
can be generated through the unitary transformation.
As for the treatment of the spurious center-of-mass (c.m.) motion,
we follow the Gloeckner-Lawson prescription~\cite{Gloeckner74}
as has been done in many of the shell-model calculations in which
a multi-$\hbar \Omega$ model space is considered.
Thus, the present shell-model Hamiltonian is composed of the
transformed Hamiltonian which contains the sums of the kinetic energy
and the two-body effective interaction, and the c.m. Hamiltonian
which is multiplied by a large value $\beta _{\rm c.m.}$.

In the following calculations, the CD-Bonn potential~\cite{Machleidt96}
is employed as the two-body bare interaction,
and the Coulomb force is also included for the proton-proton channel.
Since we adopt the above-mentioned approximations,
our calculated results may have an $\hbar \Omega$ dependence.
We have confirmed that the $\hbar \Omega$ dependences of calculated
energy levels of $^{14-18}$C become weak in the range between
$\hbar \Omega = 14$ and $16$ MeV, 
and there appear energy minima in this region.
Thus, in this study, we use the value of $\hbar \Omega =15$ MeV
in all the calculations.
Since we introduce a minimal refinement of one-body energies
in a simple way which will be explained later,
we need to use a common value of $\hbar \Omega$.

\begin{figure}[t]
  \epsfxsize = 12.5cm
  \centerline{\epsfbox{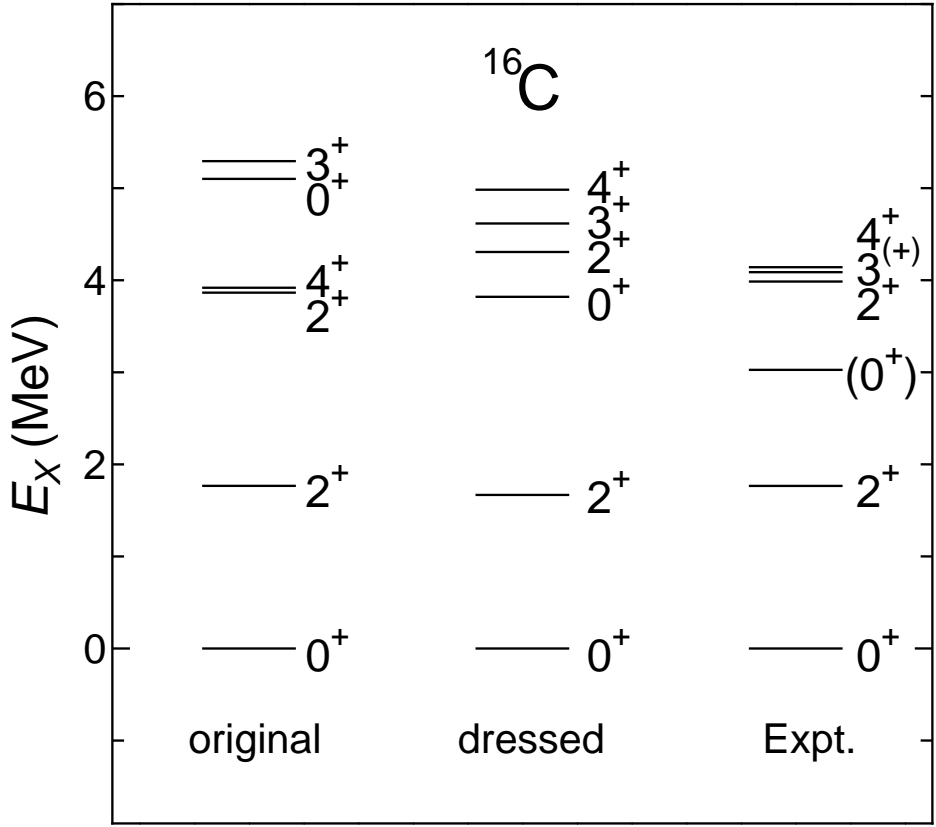}}
\caption{\label{fig:c16} Low-lying energy levels of $^{16}$C.}
\end{figure}

In Fig.~\ref{fig:c16}, calculated and experimental low-lying energy
levels of $^{16}$C are shown.
The values are also tabulated in Table~\ref{tab:c16}.
The first column in Fig.~\ref{fig:c16}, denoted by
``original", represents the results obtained directly from the
present method without any adjustable parameters.
It is seen that the calculated $2_{1}^{+}$, $2_{2}^{+}$, and
$4_{1}^{+}$ states are in good agreement with the experiment
though the $0_{2}^{+}$ and $3_{1}^{+}$ states appear somewhat higher
than the experiment.

\begin{table}[t]
\caption{\label{tab:c16} The calculated and experimental excitation
energies of low-lying excited states in $^{16}$C.
All energies are in MeV.}
\begin{center}
    \begin{tabular}{cccc}
\hline
     $J^{\pi}$ & Original & Dressed &  Expt. \\
\hline
   $2_{1}^{+}$ &    $1.77$    &  $1.67$   & $1.77$ \\
   $0_{2}^{+}$ &    $5.10$    &  $3.82$   & $3.03$ \\
   $2_{2}^{+}$ &    $3.87$    &  $4.31$   & $3.99$ \\
   $3_{1}^{+}$ &    $5.29$    &  $4.62$   & $4.09$ \\
   $4_{1}^{+}$ &    $3.92$    &  $4.98$   & $4.14$ \\
\hline
    \end{tabular}
\end{center}
\end{table}

We note here that there is a difference in the model space between
the present approach and the UMOA to neutron-rich carbon isotopes.
In the present shell-model calculation for $^{15}$C,
the $1/2^{+}_{1}$ state appears above the $5/2^{+}_{1}$ state,
contrary to the experiment.
This wrong ordering may be one of the reasons for the discrepancy
between the results and the experimental values in the excited states
of $^{16}$C for the ``original" case in Fig.~\ref{fig:c16}.
If we consider a sufficiently large model space and calculate
these energy levels within the UMOA framework,
which is fully microscopic, using the same CD-Bonn potential,
we can reproduce the correct ordering in $^{15}$C as has been shown
in Ref.~\cite{Fujii05}.
This type of calculation is feasible for describing single-particle
levels of one-nucleon plus closed-shell nuclear systems,
but becomes extremely complicated for the excited states of $^{16}$C.
In order to obtain the correct ordering in $^{15}$C in the present
shell model, we make a minimal refinement of neutron one-body
energies on top of $^{14}$C
so as to reproduce the UMOA results for $^{15}$C~\cite{Fujii05}.
In this way, we treat complex correlations coupled to a
single-particle-like state within a rather simple framework.
Thus, the present approach is a hybrid method combining a no-core
type of shell model with single-particle information by the UMOA.
Hereafter, we refer to the calculated results with the minimal
refinement as ``dressed".

The magnitude of the refinement may depend on the orbit.
In fact, denoting the one-body energy for a single-particle orbit $j$
of the neutron as $\epsilon ^{n}(j)$,
we need the variations as
$\Delta \epsilon ^{n}(1s_{1/2})=-2.18$ and
$\Delta \epsilon ^{n}(0d_{5/2})=-0.04$ MeV which are added to
the original $\epsilon ^{n}({j})$'s, namely, the
h.o. kinetic energies.
We also change the one-body energy for the $0d_{3/2}$ orbit of
the neutron by the same amount as $\Delta \epsilon ^{n}(0d_{5/2})$
for simplicity.
Since the variations of $\Delta \epsilon ^{n}(0d_{5/2})$ and
$\Delta \epsilon ^{n}(0d_{3/2})$ are very small, the effect of the
refinement arises mainly from only $\Delta \epsilon ^{n}(1s_{1/2})$.
We use these dressed one-body energies not only in the calculation
for $^{15}$C but also $^{14}$C, $^{16}$C, and $^{18}$C.
A similar approach may be useful, in particular, to some exotic systems
for which single-particle levels are unknown experimentally.

The results of the low-lying energy levels of $^{16}$C with
the above-mentioned refinement are shown in Fig.~\ref{fig:c16} and
Table~\ref{tab:c16} with ``dressed".
The calculated results become better by introducing
the dressed one-body energies.
Namely, the correct ordering of the low-lying energy levels is
obtained.
It is noted here that a recent NCSM calculation for
$^{16}$C with the CD-Bonn potential in a $4$ $\hbar \Omega$ space
also gives the correct ordering and the calculated
$B$($E2; 2_{1}^{+} \rightarrow 0_{1}^{+}$) is about $1$
$e^{2}$fm$^{4}$ which is not so different form our values in
Table~\ref{tab:be2}~\cite{Vary06}.

\begin{table}[t]
\caption{\label{tab:be2} The calculated and experimental
$B$($E2; 2_{1}^{+} \rightarrow 0_{1}^{+}$) values in units of
$e^{2}$fm$^{4}$
 for $^{14,16,18}$C.
The experimental value for $^{16}$C is taken from Ref.~\cite{Imai04}.}
\begin{center}
    \begin{tabular}{cccc}
\hline
      Isotopes &  Original & Dressed &  Expt.\\
\hline
      $^{14}$C  &  $3.42$  &  $3.42$  & $3.74\pm 0.50$ \\
      $^{16}$C  &  $1.30$  &  $0.84$  &
      $0.63^{\pm 0.11({\rm stat})}_{\pm 0.16({\rm syst})}$ \\
      $^{18}$C  &  $1.19$  &  $2.10$  &        \\
\hline
    \end{tabular}
\end{center}
\end{table}

\begin{table}[t]
\caption{\label{tab:decomposition} The decomposition of the
calculated $B$($E2; 2_{1}^{+} \rightarrow 0_{1}^{+}$) values
in units of $e^{2}$fm$^{4}$ for $^{14,16,18}$C in the case of
``dressed".}
\begin{center}
    \begin{tabular}{cccc}
\hline
      Isotopes &  Up to 0$p$ & Up to 1$s$0$d$ & Full \\
\hline
      $^{14}$C  &  $4.11$  &  $4.27$  & $3.42$ \\
      $^{16}$C  &  $0.47$  &  $0.59$  & $0.84$ \\
      $^{18}$C  &  $1.74$  &  $1.99$  & $2.10$ \\
\hline
    \end{tabular}
\end{center}
\end{table}

In Table~\ref{tab:be2}, the calculated and experimental values of
$B$($E2; 2_{1}^{+} \rightarrow 0_{1}^{+}$) for $^{16}$C are tabulated
together with the values for $^{14}$C and $^{18}$C to investigate the
isotope dependence.
For $^{16}$C, the calculated $B$($E2$) value becomes closer to
the anomalously hindered experimental value by introducing the dressed
one-body energies.
As we have explained before, in the present calculation, we do not
employ the effective charges for the neutron and proton but use
the bare charges.

For $^{14}$C,
the calculated results of $B$($E2; 2_{1}^{+} \rightarrow 0_{1}^{+}$)
for ``original" and ``dressed" are identical, and show a good
agreement with the experiment.
We note, however, that the calculated energy spacings between
the $0^{+}_{1}$ and $2^{+}_{1}$ states for the cases ``original" and
``dressed" are identically $5.39$ MeV,
whereas the experimental value is $7.01$ MeV.
Thus, the calculated energy spacings are somewhat smaller than
the experiment.
The energy spacing is sensitive to the magnitude of the spin-orbit
splitting for the 0$p$ state of the proton.
The calculated small energy spacing indicates that the proton
spin-orbit splitting in the present calculation is not sufficient
for the energy spacing.
The proton spin-orbit splitting may be larger if we enlarge the
model space and/or include a genuine three-body force in the
present approach.

For $^{18}$C, the calculated energy spacings between the $0_{1}^{+}$
and $2_{1}^{+}$ states for ``original" and
``dressed" are $1.44$ and $1.73$ MeV, respectively.
Since the experimental value is $1.62$ MeV, our calculated results are
not so different from the experiment.
As for the $B$($E2; 2_{1}^{+} \rightarrow 0_{1}^{+}$),
as shown in Table~\ref{tab:be2}, the result for ``dressed" is
larger than that for ``original".
Furthermore, the result for $^{18}$C for ``dressed" is about three
times larger than the experimental value for $^{16}$C.
A similar tendency is seen in the 0$p$1$s$0$d$ shell-model
calculations~\cite{Fujimoto03,Sagawa04}.
Thus, it is of great interest that the $B$($E2$) value for $^{18}$C
is experimentally established.

In order to investigate the $B$($E2$) results in more detail,
for the case ``dressed",
we have decomposed the $B$($E2$) values with the restrictions
up to the 0$p$ or the 1$s$0$d$ shell using the full wave functions.
The results are tabulated in Table~\ref{tab:decomposition}.
For $^{16}$C, about a half of the total $B$($E2$) is obtained
in the result up to the 0$p$ shell, and a large contribution from
the 1$p$0$f$ shell can be observed though the magnitude itself is
small.
For $^{18}$C, most of the contribution to the total $B$($E2$) is
from the orbits up to the 0$p$ shell.
For $^{14}$C, a reduction of the $B$($E2$) is seen by including
the 1$p$0$f$ shell.
Note that these results come from only the proton contributions
because we use the bare charges.

\begin{figure}[t]
  \epsfxsize = 12.5cm
  \centerline{\epsfbox{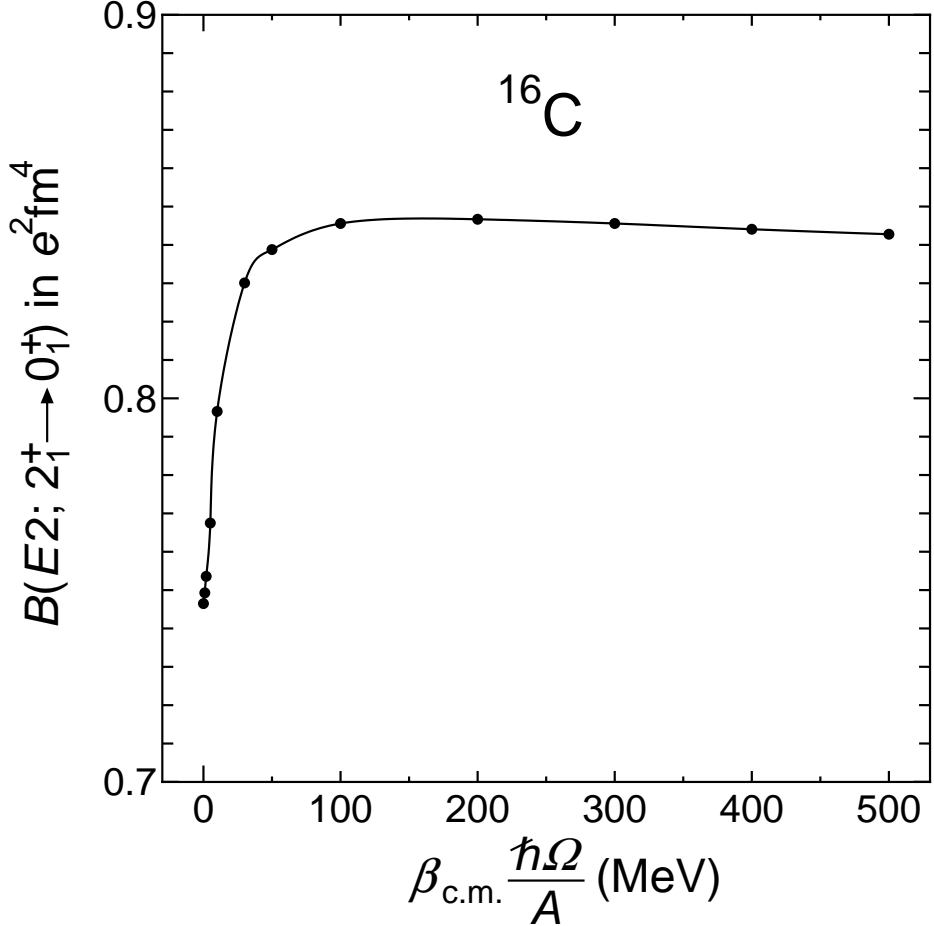}}
\caption{\label{fig:beta_be2} The $\beta _{\rm c.m.}$ dependence
of the calculated $B$($E2; 2_{1}^{+} \rightarrow 0_{1}^{+}$) values
for $^{16}$C for the case ``dressed".}
\end{figure}

Here we examine the dependence of calculated results on
$\beta _{\rm c.m.}$.
As we have explained before, we employed the Gloeckner-Lawson
prescription to remove the spurious c.m. motion.
Unlike the usual NCSM, since the present shell-model space is
composed of an incomplete space in energy, spurious c.m. components
are not removed completely.
Therefore, our calculated results have some $\beta _{\rm c.m.}$
dependence.
However, as for the $B$($E2; 2_{1}^{+} \rightarrow 0_{1}^{+}$)
for $^{16}$C which is the main interest in the present study, 
the $\beta _{\rm c.m.}$ dependence is quite small if we take
a sufficiently large value of $\beta _{\rm c.m.}$.
The tendency is illustrated in Fig.~\ref{fig:beta_be2} for the
case ``dressed".
In a wide range between
$\beta _{\rm c.m.}\hbar \Omega /A=50$ and
$500$ MeV, the differences are within
$0.01$ $e^{2}$fm$^{4}$.
We have also confirmed that the $\beta _{\rm c.m.}$ dependence of
the calculated energy spacing between the $2^{+}_{1}$ and
$0^{+}_{1}$ states is also quite small in the same range.
The differences are within $0.02$ MeV.
Therefore, our results for $^{16}$C will be reliable for the
removal of the spurious c.m. motion.
However, we should note that, for $^{14}$C,
the situation becomes worse.
For $\beta _{\rm c.m.}\hbar \Omega /A=50$ and $500$ MeV,
the calculated energy spacings for the case ``dressed" are
$5.39$ and $5.11$ MeV, respectively.
Gradually the energy spacing becomes smaller as the $\beta _{\rm c.m.}$
value is larger, and the $\beta _{\rm c.m.}$ convergence is slow.
Therefore, we will still have room to improve the method of
the removal of the spurious c.m. motion for $^{14}$C.
Furthermore, we should include many-particle many-hole excitations
higher than the $2p2h$ excitation for the realistic description of
excited states in $^{14}$C higher than $6$ MeV excitation energy.
These would apply to lighter carbon isotopes.
For these reasons, we have not performed detailed calculations for
$^{14}$C and lighter carbon isotopes.
It should be noted, however, that the situation is different
in $^{16}$C and $^{18}$C.
Since we describe the excited states below $6$ MeV excitation
energy where neutron excitations within the 1$s$0$d$ shell play
an important role,
the approximations in the present calculation may be good for $^{16}$C
and $^{18}$C.
In the present study, we used the common value
of $\beta _{\rm c.m.}\hbar \Omega /A=50$ MeV in the final results for
$^{14}$C, $^{16}$C, and $^{18}$C.

One may be interested in the calculated ground-state energy
since the present shell model is a kind of no-core calculation. 
For $^{14}$C, our calculated result for the case ``dressed"
is $-80.18$ MeV.
If we include $4p4h$ excited configurations, the result becomes
$-90.90$ MeV.
(Note that the inclusion of $4p4h$ configurations in the calculations
for $^{16}$C and $^{18}$C is not so easy due to a huge computation
size.)
We notice that since the present shell model is not
fully microscopic and needs some approximations,
the above value is not the exact one.
On the other hand, we can microscopically calculate the ground-state
energy of the closed-shell nucleus $^{14}$C within the UMOA framework.
In the case of a sufficiently large value of $\rho _{1}$,
the calculated result is $-87.70$ MeV without the three-or-more-body
cluster effects.
The inclusion of three-body-cluster terms gives an additional
attractive energy about $-3$ MeV.
Therefore, the present shell-model result with
the $4p4h$ effect is comparable with the UMOA result.
Since the experimental value is $-105.28$ MeV, the remaining attractive
energy would be brought by including a genuine three-body force.

\begin{table}[t]
\caption{\label{tab:c16_occupation} The occupation numbers
for the 0$p_{3/2}$, 0$p_{1/2}$, 1$s_{1/2}$, and 0$d_{5/2}$ orbits of
the proton ($p$) and neutron ($n$) for the $0_{1}^{+}$ and $2_{1}^{+}$
states in $^{14}$C and $^{16}$C for the case ``dressed".}
\begin{center}
    \begin{tabular}{ccccccc}
\hline
   Isotopes &  $J^{\pi}$  & Nucleon &$0p_{3/2}$&$0p_{1/2}$&$1s_{1/2}$&
$0d_{5/2}$\\ \hline
   $^{14}$C & $0_{1}^{+}$ & $p$ & $3.59$ & $0.27$ & $0.09$ & $0.02$ \\
            &             & $n$ & $3.88$ & $1.94$ & $0.08$ & $0.02$ \\
            & $2_{1}^{+}$ & $p$ & $2.83$ & $1.03$ & $0.09$ & $0.02$ \\
            &             & $n$ & $3.87$ & $1.95$ & $0.07$ & $0.03$ \\
   $^{16}$C & $0_{1}^{+}$ & $p$ & $3.41$ & $0.42$ & $0.12$ & $0.03$ \\
            &             & $n$ & $3.88$ & $1.95$ & $0.97$ & $1.01$ \\
            & $2_{1}^{+}$ & $p$ & $3.39$ & $0.44$ & $0.11$ & $0.03$ \\
            &             & $n$ & $3.88$ & $1.94$ & $0.81$ & $1.17$ \\
\hline
    \end{tabular}
\end{center}
\end{table}

In Table~\ref{tab:c16_occupation}, we show calculated occupation
numbers for the $0_{1}^{+}$ and $2_{1}^{+}$ states in $^{14}$C and
$^{16}$C for the case ``dressed".
The occupation numbers of the proton for $^{16}$C are hardly changed
between the $0_{1}^{+}$ and $2_{1}^{+}$ states.
As for the neutron, only the occupation numbers for the 1$s_{1/2}$ and
0$d_{5/2}$ orbits are slightly different between the $0_{1}^{+}$ and
$2_{1}^{+}$ states.
This result supports the decoupling of the valence two neutrons from
the core part~\cite{Elekes04,Ong06,Fujimoto03}.
In the case of $^{14}$C, the proton occupation numbers for the
0$p_{3/2}$ and 0$p_{1/2}$ orbits are rather different between the
$0_{1}^{+}$ and $2_{1}^{+}$ states in contrast to the case of $^{16}$C.
These results indicate that the excitation mechanism is very different
between $^{14}$C and $^{16}$C.

Here we comment on the ratio of the neutron and proton transition
matrix elements for $^{16}$C.
Large ratios have been found through experiments as 
$M_{n}/M_{p}=7.6\pm 1.7$~\cite{Elekes04} and
$6.7\pm 1.3$~\cite{Ong06}.
In our calculation for the case ``dressed", the ratio is $4.68$
and somewhat smaller than the experiments.
Our model space $P_{2}$ may be large enough for $M_{p}$ because of
the inclusion of the $2\hbar \Omega$ excited 1$p$0$f$ shell
from the 0$p$ shell, while it may not be sufficient
for $M_{n}$ due to the missing $2\hbar \Omega$ excited 2$s$1$d$0$g$
shell from the 1$s$0$d$ shell.
Thus, $M_{n}$ can be enlarged to a certain extent in the calculation
with a larger model space.

\begin{figure}[t]
  \epsfxsize = 12.5cm
  \centerline{\epsfbox{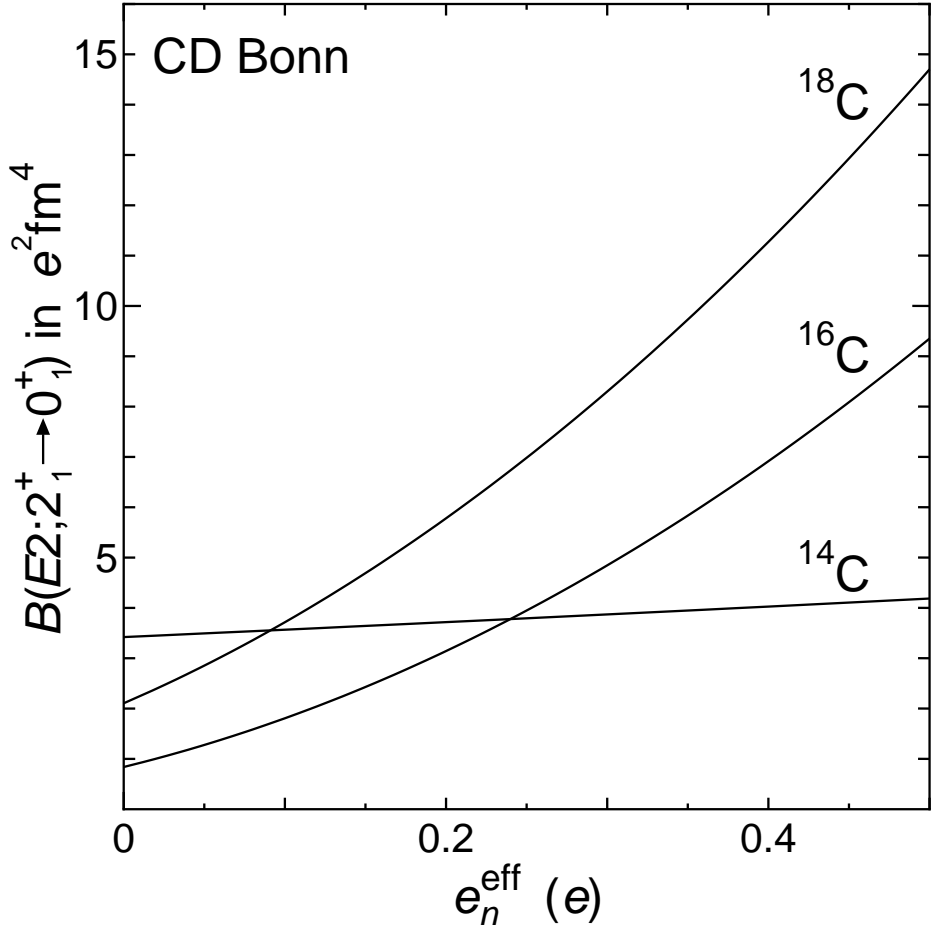}}
\caption{\label{fig:be2_echarge} The dependence of the calculated
$B$($E2; 2_{1}^{+} \rightarrow 0_{1}^{+}$) values on the neutron
effective charge $e_{n}^{\rm eff}$
for $^{14,16,18}$C for the case ``dressed".}
\end{figure}

The large $M_{n}/M_{p}$ ratio means that the transition
matrix element for the neutron $M_{n}$ is considerably larger than
that for the proton $M_{p}$.
This also indicates that the $B$($E2; 2_{1}^{+} \rightarrow 0_{1}^{+}$)
of $^{16}$C is sensitive to the value of the neutron effective charge.
This situation is illustrated in Fig.~\ref{fig:be2_echarge}.
The results for $^{14}$C and $^{18}$C are also shown.
For $^{16}$C, the $B$($E2$) value is rapidly increasing as the neutron
effective charge $e^{\rm eff}_{n}$ becomes larger.
For $e^{\rm eff}_{n}=0.5e$ which is the standard value in the
conventional shell model, the calculated $B$($E2$) is $9.35$
$e^{2}$fm$^{4}$.
This value is about ten times larger than the result for the bare
charge, namely, $e^{\rm eff}_{n}=0$.
Also for $^{18}$C, a similar tendency of the strong dependence on
the neutron effective charge is observed, while the neutron effective
charge dependence for $^{14}$C is very weak.

Although it seems that the use of the bare charges in the calculation
of the $B$($E2$) in $^{14}$C and $^{16}$C is a good approximation,
the situation is different for $^{15}$C.
The calculated $B$($E2; 5/2_{1}^{+} \rightarrow 1/2_{1}^{+}$) of
$^{15}$C with the bare charges is $0.28$ $e^{2}$fm$^{4}$.
If we make the same analysis of
the $B$($E2; 5/2_{1}^{+} \rightarrow 1/2_{1}^{+}$) in $^{15}$C as in
Table~\ref{tab:decomposition}, results for $^{15}$C are $0.15$ and
$0.19$ $e^{2}$fm$^{4}$ for the cases ``up to 0$p$" and
``up to 1$s$0$d$", respectively.
It is seen that the effect of the 1$p$0$f$ shell has a sizeable
contribution to the total $B$($E2$) as in the case of $^{16}$C.
The experimental $B$($E2$) of $^{15}$C has been determined with
very high accuracy as
$B$($E2; 5/2_{1}^{+} \rightarrow 1/2_{1}^{+})=0.97\pm 0.02$
$e^{2}$fm$^{4}$~\cite{c15_be2}.
Thus, our calculated result is considerably smaller than the
experiment.
This may indicate that the present approach is not sufficient for
describing the structure of $^{15}$C, as we introduce the minimal
refinement for the energy levels of $^{15}$C.
However, we have found that the discrepancy of the $B$($E2$) in
$^{15}$C can be compensated by introducing the neutron effective charge
as $e^{\rm eff}_{n}=0.164e$ which is much smaller than the usual
neutron effective charge in the conventional shell model.
The need of the effective charge in the structure
calculations for $^{15}$C and $^{16}$C has also been discussed
in $^{14}{\rm C}+n$ and $^{14}{\rm C}+n+n$
models~\cite{Suzuki04,Horiuchi06,Hagino07}.

If we calculate the $B$($E2; 2_{1}^{+} \rightarrow 0_{1}^{+}$) values
using the neutron effective charge $e^{\rm eff}_{n}=0.164e$ for
$^{14,16,18}$C, the results become $3.66$, $2.62$, and $4.98$
$e^{2}$fm$^{4}$, respectively.
Due to the strong neutron effective charge dependence of $B$($E2$)
in $^{16}$C and $^{18}$C as shown in Fig.~\ref{fig:be2_echarge},
the results for $^{16}$C and $^{18}$C are significantly larger than
those for the bare charges despite the rather small effective
charge of the neutron.
Therefore, within the present shell-model framework, we cannot describe
properly the $B$($E2$) values of $^{15}$C and $^{16}$C simultaneously,
as long as we employ the same value of the neutron effective charge.
This problem may be resolved if we enlarge the model space and/or
include a genuine three-body force in the present method.

In summary, we have investigated low-lying energy levels
and $B$($E2; 2_{1}^{+} \rightarrow 0_{1}^{+}$) of $^{16}$C by
introducing a new shell-model method of no-core type with the model
space up to the 1$p$0$f$ shell.
The two-body effective interaction is microscopically derived
from the CD-Bonn potential and the Coulomb force through
a unitary-transformation theory.
The present work is the first study which can successfully
describe the low-lying structure of $^{16}$C including its anomalously
hindered $B$($E2; 2_{1}^{+} \rightarrow 0_{1}^{+}$) value based on
the microscopic effective interaction with the bare charges.
In principle, we should use an effective operator in
the calculation of the transition strength~\cite{Stetcu05} as well as
the effective interaction for the energy.
Such a study should be pursued in the future.
However, we expect that the use of the bare charges in the present
model space is a good approximation for $^{16}$C.
There are some indications that the effective charges should be
smaller as the neutron richness increases in boron and carbon isotopes
even in the smaller 0$p$1$s$0$d$
space~\cite{Fujimoto03,Sagawa01,Ogawa03}.
In general, the magnitude of the effective charges for reproducing
the experimental value should depend on the model-space size employed
and each nucleus due to different correlations, such as different
core-polarization effects between stable and halo nuclei~\cite{Kuo97}.

In the present study, we do not include a genuine three-body force.
The recent GFMC and NCSM results indicate that the bare three-body
force mainly changes the binding energy but rather keep
the excitation spectrum unchanged,
except for a few cases such as beryllium and boron
isotopes~\cite{Pieper05},
while the three-body-force effect depends on the two-body
interaction employed in the calculation.
We expect that the present microscopic approach with the two-body
interaction is a reasonable one for the calculations of excitation
energies and transition strengths of low-lying states in nuclei around
$^{16}$C.
Although the successful description of $^{16}$C seems to support
this expectation,
the extension of the present method for the inclusion of
the effective three-body interactions from the bare two- and three-body
forces may be interesting for a more realistic description.
This sort of study may also improve the difficult situation for
the description of the structure of $^{15}$C.

We are grateful to Y. Utsuno for a useful discussion
on the treatment for the spurious c.m. motion.
We thank A. Gelberg for reading the manuscript.
This work was supported in part by a Grant-in-Aid for Specially
Promoted Research (Grant No. 13002001) and a Grant-in-Aid for Young
Scientists (B) (Grant No. 18740133) from the Ministry of Education,
Culture, Sports, Science, and Technology in Japan.

\end{document}